\newcommand{\be}{\begin{equation}}
\newcommand{\ee}{\end{equation}}
\newcommand{\ber}{\begin{eqnarray}}
\newcommand{\eer}{\end{eqnarray}}

\newcommand{\lsim}{\raisebox{-0.7ex}{$\stackrel{\textstyle <}{\sim}$ }}
\newcommand{\thalf}{\frac{1}{2}}

\documentstyle[preprint,epsfig,epsf,aps,floats]{revtex}
\begin{document}
\tighten
\preprint{\vbox{
\hbox{DOE/ER/41132-104-INT00}}}
\bigskip
\title{First Order Kaon Condensation in Neutron Stars:  \\
Finite Size Effects in the Mixed Phase}
\author {Travis Norsen$^1$ and Sanjay Reddy$^2$}
\address{$^1$Dept. of Physics, University of Washington, Seattle, WA 98195 \\
$^2$Institute for Nuclear Theory, University of Washington, Seattle, WA 98195}
\date{\today}
\maketitle

\begin{abstract}

We study the role of Coulomb and surface effects on the phase transition from
dense nuclear matter to a mixed phase of nuclear and kaon-condensed matter.  We
calculate corrections to the bulk calculation of the equation of state (EOS)
and the critical density for the transition by solving explicitly for
spherical, cylindrical and planar structures.  The importance of Debye
screening in the determination of the charged particle profiles is studied in
some detail.  We find that the surface and Coulomb contributions to the energy
density are small but that they play an important role in the determination of
the critical pressure for the transition, as well as affecting the size and
geometry of favored structures. This changes the EOS over a wide range of
pressure and consequently increases the maximum mass by $\sim 0.1$ M$_\odot$.
Implications for transport properties of the mixed phase are also discussed.

\end{abstract}


\section {Introduction}

It is now universally thought that some type of phase transition will occur in
nuclear matter at high densities.  In particular, a transition to de-confined
quark matter must occur at sufficiently high density.  There may also exist a
transition to pion or kaon condensed matter at intermediate densities.  In any
of these cases, the phase transition is expected to soften the equation of
state (EOS) and thereby affect the mass-radius relation of neutron stars,
favoring lower maximum masses and smaller radii compared to neutron stars
containing only nucleon degrees of freedom.  Phase transitions can also
influence transport properties and weak interaction rates, which in turn affect
various potentially observable features of neutron star evolution such as its
cooling history, spin down rates, glitches and related phenomena.

A generic feature of first order phase transitions in neutron star matter is
that a mixed phase region, where the nuclear phase coexists with a denser and
energetically favorable new phase, is mechanically stable and can thereby
occupy a significant spatial region in the interior of a neutron star
\cite{G1}.  The best studied example of a first order transition at high baryon
density is the quark-hadron phase transition, where model calculations indicate
that the mixed phase is favored for a wide range of pressure and consequently
can occupy a large spatial extent in the neutron star interior (2-6 km).

In a recent article, Glendenning and Schaffner studied a model for kaon
condensation, where the transition to the kaon condensed phase is first order
\cite{GS}. Thus, unlike earlier studies of kaon condensation \cite{KN,TPL},
where the transition was second order, the model due to Glendenning and
Schaffner predicts a mixed phase where the kaon phase coexists with a nuclear
phase. This naturally leads to the question of how surface and coulomb effects
will influence the extent and structure of the mixed phase.  Earlier studies,
in the context of a quark-hadron phase transition, showed that surface and
Coulomb effects are important \cite{HPS}.  It was found that the range of
pressures over which the mixed phase is favored depended sensitively on the
(poorly known) surface tension between the nuclear and quark phase \cite{HPS}.

In the present work, we calculate surface and Coulomb contributions (including
the effects of Debye screening) to the thermodynamic potential in a mixed phase
containing kaons.  We identify their importance in the determination of the
equation of state, the size and dimensionality of spatial structures, and the
critical density for the onset of the mixed phase.

We begin by describing the model for a first order kaon condensate in \S2.
After a critical review of the standard procedure for dealing with finite size
effects, the surface and coulomb contributions to the Gibbs potential are
evaluated in \S3.  \S4 addresses the changes induced by inclusion of Debye
screening of electrons.  In \S5, we discuss the implications of our findings to
the structure of neutron stars and transport properties in the mixed phase
region.  We summarize our main findings, and discuss their implications in
\S6.

\section {First order kaon condensation}
In this section we briefly review the model proposed by Glendenning and
Schaffner \cite{GS} which predicts a first order transition from the nuclear to
the kaon condensed phase. The nuclear phase is described by a relativistic,
Walecka type, field theoretical model including the isovector meson $\rho$ as
well as cubic and quadratic self-interactions of the scalar meson
\cite{SW,GBOOK}.  The Lagrangian for the nucleon sector is given by

\begin{eqnarray}
{\cal L}_N \!=&& \overline{\Psi}_N \! \left( i\gamma^\mu
\partial_\mu-m_N^\ast
-g_{\omega N}\gamma^\mu V_\mu -g_{\rho N}\gamma^\mu
\vec{\tau}_N\cdot \vec{R}_\mu \!\right)\! \Psi_N \nonumber
\\
&&{} +\frac{1}{2}\partial_\mu \sigma
\partial^\mu\sigma-\frac{1}{2}m_\sigma^2\sigma^2-U(\sigma)-\frac{1}{4}
V_{\mu\nu}V^{\mu\nu} \nonumber 
\\ 
&&{} +\frac{1}{2}m_\omega^2V_\mu
V^\mu-\frac{1}{4}\vec{R}_{\mu\nu}
\cdot\vec{R}^{\mu\nu}+\frac{1}{2}m_\rho^2\vec{R}_\mu \cdot
\vec{R}^\mu,
\end{eqnarray}
where $m_N^\ast = m_N-g_{\sigma N}\sigma$ is the nucleon effective mass, which
is reduced compared to the free nucleon mass due to the scalar field $\sigma$.
The vector fields corresponding to the omega and rho mesons are given by
$V_{\mu\nu} = \partial_\mu V_\nu - \partial_\nu V_\mu$, and $ \vec{R}_{\mu\nu}
= \partial_\mu \vec{R}_\nu -\partial_\nu \vec{R}_\mu $ respectively.  The
scalar self-interaction term is given by $U(\sigma)= (1/3)bm_N(g_{\sigma
N}\sigma)^3 + (1/4)c(g_{\sigma N}\sigma)^4$, where $b$ and $c$ are
dimensionless coupling constants.  $\Psi_N$ is the nucleon field operator with
$\vec{\tau}_N$ the nucleon isospin operator.  The five coupling constant,
$g_{\sigma N}$, $g_{\omega N}$, $g_{\rho N}$, $b$, and $c$, are chosen to
reproduce the empirical properties of nuclear matter at saturation density
\cite{SW}.

Kaons are included in the model in the same fashion as the nucleons, by
coupling to the $\sigma$, $\omega$ and $\rho$ meson fields. There exist in the
literature several meson-exchange Lagrangians which attempt to describe
kaon-nucleon interactions. A detailed discussion of these models and their
relation to the Chiral Lagrangian proposed by Kaplan and Nelson \cite{KN} 
can be
found in Pons, et al. \cite{PRPPL}.  In the present paper we employ the
Lagrangian proposed by Glendenning and Schaffner. The kaon Lagrangian is 
given by
\begin{equation}
{\cal L}_K ={\cal D}_\mu^\ast K^\ast {\cal D}^\mu K
-m_K^{\ast 2} K^\ast K, 
\end{equation}
where $K$ denotes the isospin doublet kaon field.  The covariant derivative $
{\cal D}_\mu = \partial_\mu+ig_{\omega K} V_\mu+ig_{\rho K} \vec{\tau}_K \cdot
\vec{R}_\mu $ couples the kaon field to the vector mesons and the kaon
effective mass term $m_K^\ast = m_K-g_{\sigma K}\sigma$ describes its coupling
to the scalar meson. $\vec{\tau}_K$ is the kaon isospin operator.  The vector
coupling constants are determined by isospin and quark counting rules
\cite{GS}, and are given by $g_{\omega K}=g_{\omega N}/3$ and $g_{\rho
K}=g_{\rho N}$. The scalar coupling is fixed by fitting to an empirically
determined kaon optical potential in nuclear matter. Albeit poorly known, the
real part of the optical potential in nuclear matter as deduced from kaonic
atoms indicates that 80 MeV $ \lsim U_K(n_o) \lsim $180 MeV \cite{FGB,OR}.
Here, we choose $U_K(n_o)= 120$ MeV to fix $g_{\sigma K}$. Lower values of the
optical potential favor a second order transition, while higher values make
the first order transition stronger. For the purpose of this study, we require
that the phase transition be first order but the qualitative features of the
results found here are independent of its precise value.

In the mean-field approximation only the time component of the vector fields
and the isospin 3-component of the isovector field can have nonzero mean
values.  The equations of motion (EOM) for the non-strange meson fields can be
derived from the above Lagrangians and are given by
\ber
-\nabla^2 \sigma + m^2_\sigma \sigma &=& -\frac{dU}{d\sigma} + g_{\sigma
B}(\rho^s_n +\rho^s_p) + g_{\sigma k} f_\pi^2 \theta^2 m_k^* \, ,\\ 
-\nabla^2  \omega + m_\omega^2 \omega &=& g_\omega (\rho_n+\rho_p) 
- g_{\omega K} f^2_\pi \theta^2 (\mu_K+X) \, ,\\ 
-\nabla^2 r + m^2_\rho  r &=&  g_\rho (\rho_p-\rho_n) - g_{\rho K}
f^2_\pi \theta^2 (\mu_K+X) \, .
\eer
where the meson fields $\sigma, \omega, r$ now represent the appropriate mean
values, $X=g_{\omega K} \omega + g_{\rho K} r$, and $\rho ^ {(s)} _ {(n,p)}$
are the (scalar) densities of neutrons and protons.  The ansatz $K=f_K \theta /
\sqrt{2}$ has been made, where $f_K$ is the kaon decay constant and $\theta$ is
a dimensionless kaon field strength parameter.  In the bulk approximation, the
gradient terms are omitted and the EOM become non-linear algebraic equations to
be solved self-consistently for the meson fields.  The EOM for the kaon is
\begin{equation}
\left( {\cal D}_\mu{\cal D}^\mu+m_K^{\ast 2} \right)K=0.
\end{equation}
where ${m_K}^{\ast} = m_K - g_{\sigma K} \sigma$ is the kaon effective mass and
${\mu_K}^{\ast} = \mu _e+ X$. In terms of $\theta$, the kaon EOM is re-expressed
as
\be
\label{keom}
\nabla^2 \theta = \left( {m_K}^{\ast 2}-{\mu_K}^{\ast 2} \right)\theta.
\ee
In the bulk mean field calculation, the solution to Eq.(\ref{keom}) and the
other meson field equations are obtained with the gradient terms set equal to
zero. This determines the condensate amplitude and the expectation values of
the other meson fields \cite{GS} and completely specifies the ground state of
matter at the mean field level. However, surface effects are ignored since the
gradient terms in the EOM are neglected. In this work we explicitly retain the
gradient term and solve Eq.(\ref{keom}) for the kaon condensate in the
Wigner-Seitz approximation using spherical, cylindrical, or planar geometry as
appropriate. This is described in detail in \S3.

The energy density of the nuclear phase is
\ber
\label{eden}
\epsilon_N &=& \int\limits_{0}^{{k_{F_n}}}
\frac{\mathrm{d}^3k}{(2\pi)^3} \sqrt{k^2+{m_N}^{\ast 2}}
+\int\limits_{0}^{{k_{F_p}}}
\frac{\mathrm{d}^3k}{(2\pi)^3} \sqrt{k^2+{m_N}^{\ast 2}} 
+\frac{\mu^4_e}{4 \pi^2}
\nonumber \\
&+& \thalf\left( 
(\nabla\sigma)^2+{m_{\sigma}}^2 \sigma^2 
\right) 
+ \thalf \left( (\nabla\omega)^2+{m_{\omega}}^2 \omega^2\right) + \thalf \left(
(\nabla r)^2+{m_{\rho}}^2 r^2 \right) + U(\sigma)
\label{enucleon}
\eer
where ${k_F}^{(n,p)}$ are the neutron and proton Fermi momenta and $\sigma$,
$\omega$ and $\rho$ are the expectation values of the meson fields in the
absence of kaons (note the gradient terms are zero in the bulk calculation and
are included here for completeness). In the phase containing kaons, the meson
field equations are modified and consequently their expectation values will
change. Therefore, the energy density in the kaon phase due to the nucleon is
given by Eq. (\ref{enucleon}), but with different expectation values for
$\sigma$, $\omega$ and $\rho$ meson fields.  The contribution to the energy
density from the kaon field itself is
\be 
\epsilon_{\mathrm{kaon}} =
\thalf f_K^2 \left[ (\nabla\theta)^2 + (m_K^*)^2 \theta ^2
\right]  \,.
\ee
In the absence of gradient terms the kaons do not directly contribute to the
pressure as they form an s-wave condensate. However, as is well known, the
attractive interaction between nucleons and kaons lowers the nuclear 
contribution to the pressure, which is given by
\ber
\label{press}
p &=& \frac{1}{3\pi^2}\int\limits_{0}^{{k_{F_n}}}
\frac{k^4\mathrm{d}k}{\sqrt{k^2+{m_N}^{\ast 2}}}
      +\frac{1}{3\pi^2}\int\limits_{0}^{{k_{F_p}}}
\frac{k^4\mathrm{d}k}{\sqrt{k^2+{m_N}^{\ast 2}}} 
+\frac{\mu^4_e}{12 \pi^2}
\nonumber \\
&+& \thalf \left( (m_{\omega}\omega)^2 + (m_{\rho}\rho)^2 
   - (m_{\sigma}\sigma)^2 \right) - U(\sigma)
\eer
where the meson mean field values are affected by the presence of a 
kaon condensate. 

\section {Surface and coulomb effects in the mixed phase}

A complete description of the mixed phase should include surface and Coulomb
contributions to the energy density. The Coulomb energy arises because each of
the two phases is electrically charged. If the electric charge densities were
small, the size of energetically favored structures would be very large, and
the contribution of the surface tension to the overall energy density would
become vanishingly small.  In our case, however, the electric charge densities
are large (comparable to the baryon number densities), resulting in small
structures (5-7 fm) for which the surface to volume ratio is large.
Consequently the surface tension makes an important contribution to the
total energy density.

Unlike in the quark-hadron transition, where the surface tension is poorly
known, the surface tension between the kaon and nucleon phases can be
calculated within the purview of the model described earlier because the two
phases (and hence also the surface between them) are considered within the same
theoretical framework.  Before detailing the procedure we adopt in this work,
we will briefly review the methodology often employed in calculating surface
and Coulomb contributions to the thermodynamics.

\subsection{Critique of the Bulk Calculation} 

The bulk calculation (what we refer to as the ``bulk calculation" is defined
below) of droplet phase thermodynamics, especially in the context of the mixed
phase containing nuclei and dripped neutrons, is described in detail in several
earlier works \cite{RPW,PR}. In a recent work, Christiansen, {\it et al.}
\cite{CGS} have studied the kaon-nucleon mixed phase using similar methods.
The bulk calculation proceeds as follows: At a given baryon chemical potential,
the electron chemical potential is adjusted so as to obtain equal pressures in
the two phases. This determines the composition of each phase, allowing one to
calculate the volume fraction of each phase based on the requirement of global
electric charge neutrality. The various thermodynamic quantities are then
determined by volume-fraction-weighted averaging.

The bulk energy density (Eq. \ref{eden}) is supplemented with the inclusion of
surface and Coulomb terms.  The surface tension coefficient $\sigma_0$ is
estimated by computing the interface energy between two semi-infinite slabs by
explicitly including the gradient terms, as was done for the nuclear-kaon phase
boundary in Ref. \cite{CGS}. Once the surface tension is known, the surface
contribution to the energy density for a structure of size $R$ and Wigner-Seitz
radius $R_{WS}$ can be approximated by
\be
\epsilon_{Surface} = \frac{\chi \sigma_0 d}{R}
\ee
where $\chi=(R/R_{WS})^d$ is the volume fraction of the kaon phase and $d$
is the
dimensionality ($d=3$ for droplets, $d=2$ for cylinders, and $d=1$ for slabs).
The Coulomb energy is
\be
\epsilon_{Coulomb} = 2 \pi  ({\Delta \rho})^2 r^2 \chi f_d(\chi)
\ee
where
\be
f_d(\chi)=\frac{2/(d-2)(1-\frac{1}{2} d \chi^{1-2/d})+\chi}{d+2}\; ,
\ee
$\Delta \rho$ is the difference in electric charge densities of the two
phases, and $\chi$ is the volume fraction of the rare phase \cite{RPW}. 

The total energy density is then given by
\be
\epsilon=\chi\epsilon_K + (1-\chi)\epsilon_N + \epsilon_{Surface} + 
\epsilon_{Coulomb}.
\ee
The stable size for geometrical structures is found by minimizing this energy
density with respect to variations in the size $R$.  In the usual method, this
procedure assumes that the two bulk energy densities $\epsilon_N$ and
$\epsilon_K$ do not vary with $R$, so that the minimization amounts to making
the sum of surface and Coulomb contributions as small as possible.  As is
well-known, this calculation establishes that the favored size is that which
permits $E_{Surface}=2E_{Coulomb}$ independent of the dimension \cite{RPW}.

The bulk calculation is premised on: (1) pressure equilibrium between the two
phases; (2) constancy of the bulk energy densities (and surface tension) as the
structure size varies; (3) the neglect of the finite thickness of the phase
boundary; and (4) the neglect of modifications to the charged particle
distributions due to electrostatic forces (Debye screening). In what follows
we address each of these assumptions and show that none of them are valid for
the case in study.

First, the assumption of pressure equality cannot hold in general. To see this,
assume we have a system that has separated into two phases, with the total
energy given by $E=E_1+E_2+E_{Surface}$, where $E_{1(2)}$ refers to the energy
in phase 1(2).  If the surface tension is not a strong function of the shape of
the phase boundary, the surface energy will be given by $E_{Surface}=\sigma_0
A$ where $A$ is the area of the boundary.  In equilibrium, the total energy
will be minimized with respect to the volume occupied by phase 1, $V_1$.  (The
boundary will move and thereby change the volumes occupied by the two phases
until the total energy is minimized.)  Differentiating the expression for the
total energy with respect to $V_1$ and setting the result to zero gives
\ber
p_1 - p_2 &=& \sigma \frac{dA}{dV_1} \,,
\eer
where $p_1=-\partial E_1/\partial V_1$ and $p_2=-\partial E_2/\partial V_2$ are
the pressures in the two phases, which are seen to differ by an amount
proportional to the surface tension and the derivative of the boundary area
with respect to the volume of the rare phase. For the case of slabs, the
indicated derivative is zero since the boundary can move back and forth without
changing its area.  For the case of spherical droplets and cylindrical rods,
however, the derivative is proportional to $R^{-1}$ and the pressure difference
between the two phases may become large for small structures. The pressure,
however, is tied to the other thermodynamic quantities through the usual
relation $\epsilon - \sum \mu n = -p$.  Consequently, the bulk energy density,
baryon number density and the chemical potentials cannot remain constant; they
change as the size is varied due to changes in the pressure of the kaon phase.
Note that a change in the bulk properties of the kaon phase for small
structures also necessitates a change in the surface tension, which depends on
the integrated gradients between the two phases.

For small structures the finite thickness of the surface region also plays an
important role.  The thickness of the boundary layer between the two phases is
of order 5 fm.  For structures of about this size, only a very small region at
the center obtains the bulk value of the kaon field, that is, the gradient
terms in the meson fields are non-zero nearly everywhere in the kaon structure.
Thus, for small structures, the assumption of zero boundary thickness grossly
overestimates the extent of the bulk kaon region.  Any attempt to produce a
structure with a radius {\it smaller} than this thickness will result in
dramatic changes to the bulk properties of the rare phase, since the fields
don't have room to achieve their usual bulk values at the center of the
droplet.  Therefore, the bulk approximation completely breaks down for
structures smaller than this size.

Thus assumptions (1-3) of the bulk calculation can only be trusted in the limit
of relatively large structures.  In order to quantify this statement, we have
constructed a sequence of kaon-condensate droplets, by explicitly solving the
kaon EOM, having different radii and at a fixed external pressure.  Various
thermodynamic quantities are then calculated and compared to results obtained
in the bulk calculation.

\begin{figure}[tbh]
\centering
{
\epsfig{figure=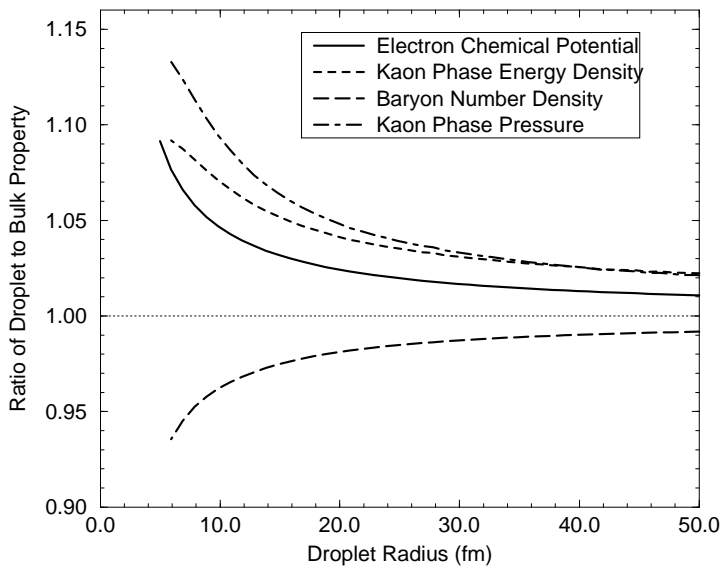, width=.65\textwidth}
}
\caption{Ratio of droplet properties to corresponding bulk properties as
a function of droplet radius. The droplets are constructed with the external
pressure (pressure in the nuclear phase) held fixed at a value 
$p = .35$fm$^{-4}$.}
\label{ratio}
\end{figure}
Fig. \ref{ratio} shows the ratio of droplet properties to bulk-calculation
properties as a function of droplet size.  As expected, the quantities
asymptote to their bulk values in the limit of large droplets, but differ from
their bulk values at the 10\% level for droplets with radii below about 10 fm.
This behavior is typical of small structures, being more pronounced for
droplets, less so for rods and is absent for slabs (for the pressure
equilibrium reasons discussed previously). These finite size effects are
therefore likely to play an important role in the vicinity of the critical
density associated with onset of the mixed phase where the favored structures
are droplets. Here, it is important to note that finite size corrections tend
to increase the kaon phase energy density and decrease the baryon number
density with decreasing radius.  This is significant because both of these
effects tend to make small droplets energetically less favorable than predicted
by the bulk calculation.  Also, the increase of the electron chemical potential
for small structures near the critical density acts to make the electric charge
density of the nuclear phase negative, making it impossible to satisfy the
charge neutrality condition, except in the case of unreasonably large
structures.  Both of these effects suggest that the critical pressure for the
transition may be increased from its naive bulk value.  This is indeed what we
find.

Thus, the first three assumptions above are valid only in the limit of large
structures.  Assumption (4), however, is known to be valid only for structures
smaller than the smallest Debye screening length. (Note that in the previous
analysis we neglected screening.)  The typical Debye screening length of
electrons at the density of interest here is 7-10 fm \cite{HPS}.  For larger
droplets the charged particle distributions will readjust to screen the droplet
charge. To demonstrate this, we show the particle density profiles for a
droplet with fixed total number of kaons, for the two cases: with and without
screening included in the equations of motion.  To include this effect, we use
a relaxation procedure in which we start with an initial guess for the shape of
the electric potential in the Wigner-Seitz cell, solve for the kaon EOM to
obtain the charged particle profiles, and then recalculate the electric
potential using the new profiles.  This is then repeated until convergence.
\begin{figure}[tbh]
\centering
{
\epsfig{figure=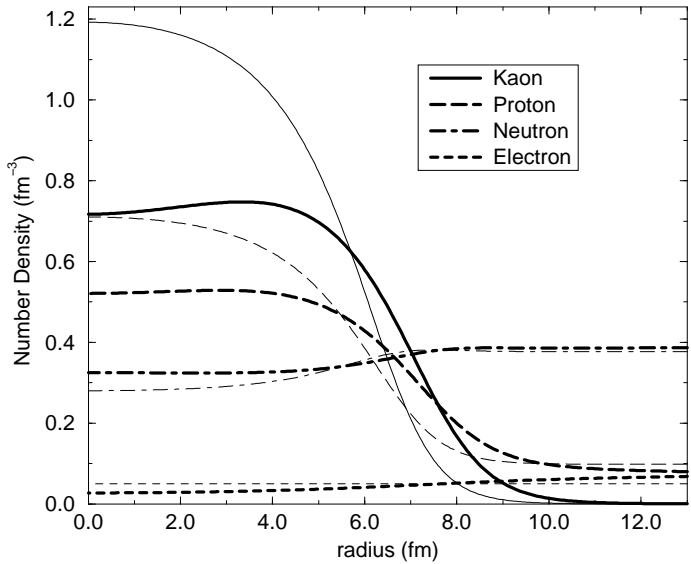, width=.65\textwidth}
}
\caption{Comparison of droplet profiles with and without electron Debye
screening.  The screened droplet (thick lines) and the unscreened droplet (thin
lines) each contain a total kaon number of $1100$.  The effect of screening is
to push many of the kaons near the surface of the droplet and thus greatly
reduce the central kaon number density.  A similar effect occurs for electrons,
although the overall densities are lower.  The neutron profile is changed only
slightly by screening, while the proton density closely follows the changes in
the kaon distribution.}
\label{profile}
\end{figure}

The effect of the electrostatic forces on the particle density profiles is
shown in Fig. \ref{profile}. The large negative potential at the center of the
droplet tends to push the negatively charged kaons toward the surface region,
and the electrons out beyond the surface. The proton distribution follows more
closely the changes in the kaon distribution since the strong interaction
between the kaons and protons overwhelms the Coulomb effect.  Screening becomes
more pronounced for droplet radii larger than typical Debye screening
length, which in the present case appears to be about 5 fm. Droplets larger
than this size are greatly deformed, with the kaons residing mainly on the
surface.

Although this deformation increases the energy density associated with the
strong interactions by forcing the fields to take on less than ideal values in
the central regions, the overall effect is to significantly reduce the
Coulomb energy of the structures.  This is essentially as expected.  After
all, allowing the fields to rearrange themselves in accordance with the
electric potential is tantamount to adding a degree of freedom to the system;
so it is no surprise that the net effect is for the system to find a way to
lower its total energy.

While the deformation due to Debye screening indeed becomes less important for
smaller structures, it appears to never become negligible.  This is also
expected, since the screening length is approximately the same size as the
boundary thickness and it is not possible for structures to become smaller than
this thickness without being greatly deformed.  The quantitative effects of
screening on the energetics and geometry of structures will be discussed
further in \S4.  For now, suffice it to say that one cannot expect {\it a
priori} the effects of Debye screening to be small compared to other finite
size effects.

To summarize, the assumptions of the bulk calculation are inconsistent.  Some
are valid only in the limit of very large structures, while the neglect of
screening is plausible only in the opposite limit. Consequently, a description
of the mixed phase calls for a more careful treatment. This is main goal of
this work.  In the what follows we describe the procedure we adopt to calculate
the thermodynamics of the mixed phase.

\subsection{Energy Minimization and Results}
Our task here is to determine the ground state configuration of a local
region of a neutron star interior at a specified pressure.  Consequently,
the appropriate thermodynamic potential is $G=E+pV$, the Gibbs
potential.  With conservation of baryon number, the
ground state will minimize the Gibbs potential per baryon:
\be 
g=\frac{G}{N_B}=\frac{E+pV}{N_B}=\frac{\epsilon+p}{n_B}\,.
\ee
Here $\epsilon$ is the energy density, $p$ is the pressure, and $n_B$ is the
baryon number density.  Global electrical neutrality is imposed by considering
the energy density and baryon number density averaged across a Wigner-Seitz
cell containing zero net charge.

\begin{figure}[tbh]
\centering
{
\epsfig{figure=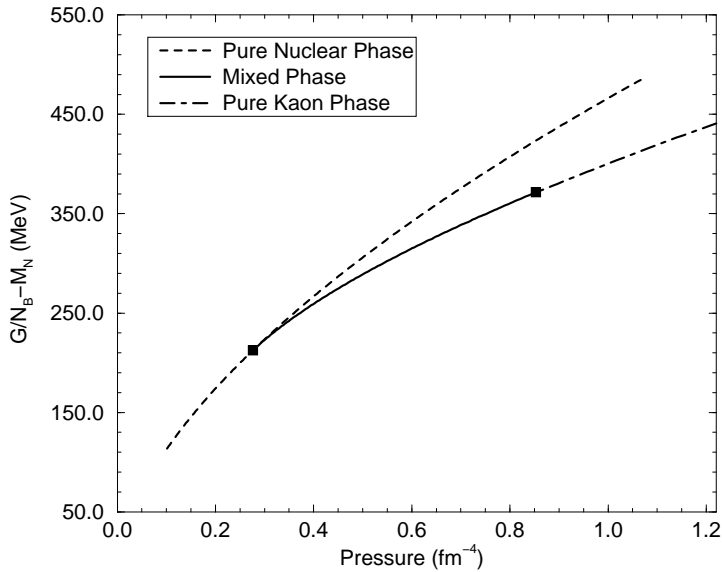, width=.65\textwidth}
}
\caption{Gibbs energy per baryon as a function of pressure for the pure
nucleon phase and the kaon mixed phase.  The mixed phase is clearly
energetically favored over the entire pressure range of its existence,
indicated by the two squares. However, the values shown here ignore surface and
Coulomb contributions to the energy as well as Debye screening. }
\label{epb}
\end{figure}

The Gibbs free energy per baryon in the kaon mixed phase, computed in the bulk
approximation is shown in Fig.\ref{epb}. The results shown do not include the
the surface and Coulomb contributions. The Gibbs energy of the homogeneous
phase containing only nucleons is also shown (solid line). The critical
pressure for the onset of the mixed phase is determined by the requirement that
the pressure in the two phases be equal, phases be oppositely charged and the
Gibbs energy be lower than that of the homogeneous nucleon phase. From the
figure we see that this occurs at a pressure $p = 0.28 $ fm$^{-4}$,
corresponding to a baryon number density of $n_B = 0.43$ fm$^{-3}$.

The goal of the present work is to identify the changes that result to this
simple picture when finite-size effects are correctly incorporated.  In order
to include surface and Coulomb contributions, we solve the kaon EOM.  We begin
with a small value of the kaon field at a finite radius, and then integrate
back toward the origin.  The chemical potentials are then adjusted until the
desired pressure is reached and the kaon field attains zero slope at the
origin.  The different geometries (spheres vs. cylinders vs. slabs) are
implemented by using the appropriate form of the Laplacian operator for that
geometry.  We omit the effect of Debye screening in what follows; it will be
discussed in detail in the next section.  The Wigner-Seitz cell size is
determined as usual by the requirement of overall charge neutrality.

The energy density is calculated as in Eq. \ref{eden} with the gradient terms
included.  This allows for exact treatment of the energy associated with the
surface and obviates the need to compute the surface tension and curvature
coefficients. The Coulomb energy density is $\epsilon_{Coulomb}=
\frac{\epsilon_0}{2}|E(r)|^2$ where $E$ is the electric field determined from
Gauss' law for the relevant geometry by explicitly integrating the electric
charge density up to a given radius.  At a given pressure, the most stable
configuration will minimize the Gibbs free energy per baryon, as described
above.  We construct a sequence of droplets/rods/slabs of varying size and
study how $G/N_B$ varies with $R$ at fixed pressure.  The results are
qualitatively as expected.  For large structures, the Coulomb energy dominates,
while for very small structures the surface energy dominates.  At intermediate
sizes there is a trade-off which results in a minimum of $G/N_B$ vs. $R$.  This
minimum determines the ground state at a given pressure.

Some details of this minimization deserve mention.  Because of the variation of
the electron chemical potential with structure size, the nucleonic matter
outside of the kaon-condensate structure sometimes becomes negatively charged
for small structures.  This occurs for pressures just above the bulk critical
pressure, where the bulk approximation predicts a nucleonic phase with small
positive charge.  If both phases become negatively charged, however, no
Wigner-Seitz cell can be defined, and the structure in question is ruled out.
Thus the requirement of overall electric charge neutrality forbids the
existence of reasonably-sized structures at pressures just above the bulk
critical pressure.

\begin{figure}[tbh]
\centering
{
\epsfig{figure=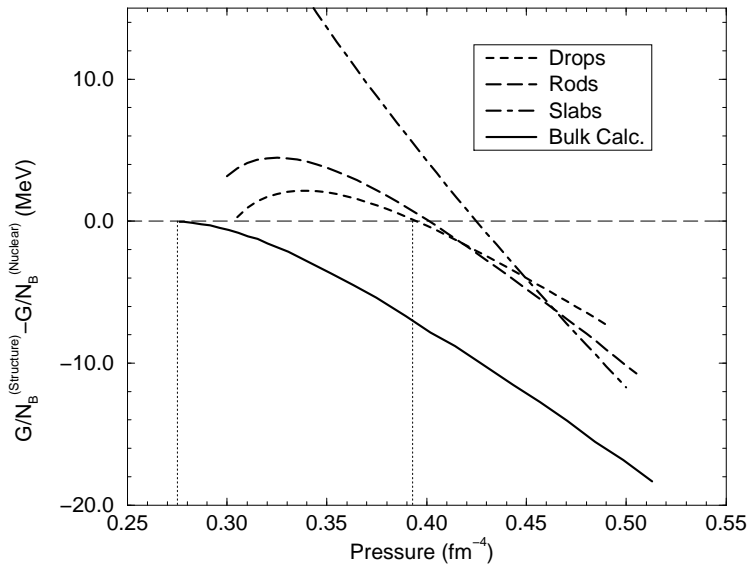, width=.65\textwidth}
}
\caption{Gibbs energy per baryon as a function of pressure for droplets,
rods, and slabs.  The energy is measured relative to the Gibbs energy per
baryon of pure nuclear matter at the same pressure; hence, a given geometric
structure is favored only when it lies below zero.  The bulk calculation
(neglecting surface and Coulomb effects) is included for comparison.  Note the
change in value of the critical pressure (i.e., the pressure marking the onset
of a mixed phase), indicated here by two vertical lines for the old and new
values.  Note also that these results omit the effects of Debye screening,
which will be discussed shortly.}
\label{gibbs}
\end{figure}

The Gibbs energy per baryon is plotted as a function of pressure for the three
different geometries in Fig. \ref{gibbs}.  Note that the values are shown
relative to the Gibbs energy of the pure nuclear phase.  As expected, there is
a smooth transition between the relative favorability of drops (which are
preferred at lower pressures) to rods and slabs as the pressure is increased.
Although we do not construct them explicitly here, this sequence is expected to
continue in the usual way through anti-rods and anti-drops in which the nuclear
phase becomes the rare phase.  The significant feature of our result is the
comparison of the structures to the Gibbs energy per baryon of electrically
neutral pure nucleon matter.  Just above the critical pressure, where droplets
are the favored geometry and therefore will be present according to the bulk
calculation, we find that the Gibbs energy is lower in the nucleonic phase than
in the mixed phase.  This remains the case over a wide range of pressures,
until droplets finally become favorable relative to pure nuclear matter at a
pressure about 50\% higher than the bulk calculation value of the critical
pressure, nearly $.40$ fm$^{-4}$ instead of $.28$ fm$^{-4}$.  At a slightly
higher pressure, rods and subsequently slabs take over as the energetically
favored geometry.

The reason for the curvature in the droplet and rod energy at low pressures
involves electric charge neutrality, as discussed previously.  At these low
pressures, the variation of the structure size necessitates a change in the
electro-chemical potential.  This change then results in the nucleonic matter
outside the kaon region becoming less and less positively charged.  At some
minimum size, the matter outside becomes uncharged, and the Wigner-Seitz radius
becomes infinite and the kaon filling fraction goes to zero.  In general, as
the structure size is reduced at constant pressure, the kaon phase volume
fraction approaches zero, and the Gibbs energy approaches (from above) its
value for pure electrically neutral nucleonic matter at the same pressure.  So
although the mixed phase energy appears to come close to the pure nucleon phase
values near the bulk critical pressure, a mixed phase is never in fact favored
at these low pressures.

The striking result that we find here is the significant increase in the
critical pressure at which the mixed phase becomes favorable.  Qualitatively,
this increase in the critical pressure should be expected since the bulk
calculation neglected the surface and Coulomb energies. An increase in the
critical pressure will make the equation of state stiffer over this range and
correspondingly the maximum mass of the neutron star will increase.  In
addition, the radial extent of the mixed phase in the interior will be
significantly reduced.

\section{Debye Screening}

The effect of Debye screening on the radial dependence of the particle
densities was shown in Fig. [\ref{profile}].  Charged particle profiles had to
readjust quite significantly to lower the coulomb energy. Clearly, the effect
of the deformation of particle profiles caused by screening is to increase the
hadronic part of the energy density associated with the structure.  This must
occur because the fields no longer take on their optimal values over a sizable
region of the structure.  The net effect of screening, however, is to lower the
total energy, since the reduction in the Coulomb energy due to screening is
larger than the cost in hadronic energy associated with the deformation. 

\begin{figure}[tbh]
\centering
{
\epsfig{figure=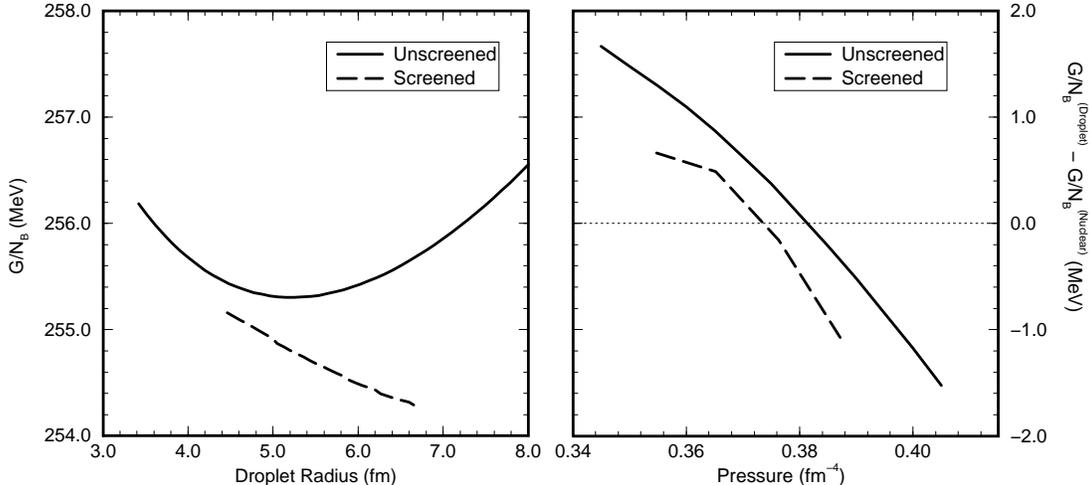, width=.85\textwidth}
}
\caption{
Gibbs energy per baryon as a function of droplet radius, for
different amounts of screening.  The screening is controlled by
shifting the electrochemical potential at each radius by a fraction
of the true electric potential.  By turning on the screening slowly,
one clearly sees the qualitative effect of the overall energy per
baryon being lowered due to the reduction in Coulomb energy.}
\label{screen}
\end{figure}

The qualitative effects of screening on the Gibbs energy are shown in
Fig. [\ref{screen}]. The results illustrate the importance of the Coulomb
energy relative to other contributions.  For example, the differences in Gibbs
energies among unscreened droplets of different sizes are small compared to the
differences between unscreened droplet and screened droplets of any
size. Screening also results in a qualitative change in the radius dependence
of the Gibbs free energy. The Gibbs free energy of the screened droplets
decreases with increasing radius for small to intermediate size droplets.

How are the results of the previous section affected by screening ?  To answer
this, we must first point out another qualitative effect of screening which
again involves electrical charge neutrality.  As shown in Fig. [\ref{profile}],
Coulomb forces tend to repel the negatively charged electrons away from the
kaon droplet.  Whenever the Wigner-Seitz radius is large (which can happen
either when the kaon structure itself is large, or when the kaon filling
fraction is low) this repulsion results in an over-abundance of electrons in
the region outside of the kaon structure.  But as before, once the electric
charge density of the nucleonic phase becomes negative, it is no longer
possible to define a Wigner-Seitz cell, and the structure under consideration
is forbidden.

We find that precisely this problem occurs for all droplets below a pressure of
about $p=.35$ fm$^{-4}$.  Above this pressure, where it is possible to
construct a fully screened globally electrically neutral droplet, the Gibbs
energy per baryon is indeed lowered somewhat from the values plotted in
Fig. [\ref{gibbs}].  This results in the critical pressure decreasing slightly
from the new value quoted earlier, but the change here is small.  The first
fully screened structure we can produce with Gibbs energy per baryon below that
of the pure nucleonic phase occurs at a pressure of $.375$ fm$^{-4}$ (compare
this with the critical pressure $\sim 0.385$ fm$^{-4}$ when Debye screening was
ignored).  Thus, while the Debye screening can have large effects on the
particle density profiles, it has a relatively small effect on the free energy
density (less than one MeV per baryon), as can be seen in the right panel of 
Fig. \ref{screen}.  Consequently
the results of the previous section remain qualitatively unchanged: the
inclusion of finite-size effects significantly increases the critical pressure
for transition to a mixed phase.

Because the changes in Coulomb energy associated with screening are large
compared to the energy differences between various geometries at a given
pressure, screening will also play a dominant role in the determination of
structures that are preferred at a given pressure.  The qualitative feature of
droplets being replaced by rods (and subsequently slabs) as the pressure is
increased is expected to remain.  But the particular pressures at which the
transition from one geometry to the next occurs may vary from the picture of
Fig. [\ref{gibbs}]. Thus, if one is only interested in the long wavelength
thermodynamic properties of the system, it is safe to ignore Debye screening.
However, to describe the mixed phase in more detail, i.e, specify the particle
profiles and the geometries that are favored at a given pressure Debye
screening must be accounted for.

\section{Neutron Star Properties}
\subsection{Structure}
\begin{figure}[tbh]
\centering
{
\epsfig{figure=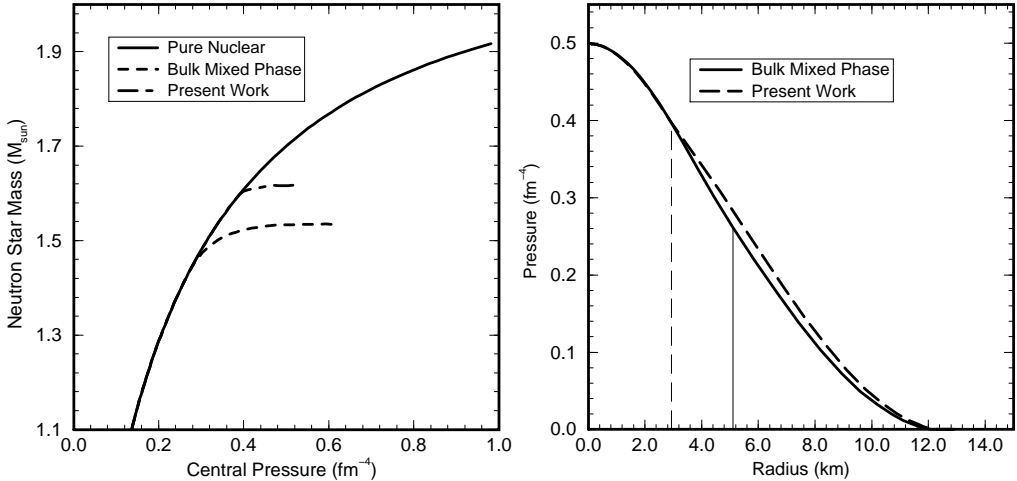, width=.85\textwidth}
}
\caption{Mass as a function of the central pressure for a pure nuclear 
equation of state, the bulk mixed calculation without finite-size effects, and
the realistic calculation of the present work.  A correct treatment of surface
and Coulomb energies increases the critical pressure for the realization of the
mixed phase, and the neutron star maximum mass by $\sim 0.1 M_{\odot}$. The
right panel compares the internal structure of a star with and without finite
size corrections included in a description of the mixed phase.  The vertical
lines indicate the outer edge of the mixed phase for the two descriptions.}
\label{ns}
\end{figure}

The equation of state determines the neutron star structure through the
Tolman-Oppenheimer-Volkov (TOV) equation:
\be
\frac{dp}{dr} = -\frac{[p(r)+\epsilon (r)][M(r)+4 \pi r^3 p(r)]}
  {r(r-2M(r))}
\ee
where $p$ is the pressure, $\epsilon$ is the energy density, and
\be
M(r)= 4 \pi \int_{0}^{r} \epsilon (r) r^2 \mathrm{d}r
\ee
is the total mass enclosed at radius $r$.  The solution to the TOV equation are
shown in Fig. \ref{ns} for three cases: pure nuclear matter, the bulk
calculation of the mixed phase, and a calculation of the mixed phase as
described in this work which properly accounts for surface and coulomb energy.
The left panel of the figure shows the variation of the total gravitational
mass as a function of the central pressure. We find that the maximum neutron
star mass is increased by about $0.1 M_{\odot}$ relative to the bulk
calculation of the mixed phase. This is as expected since the critical
pressure for the transition was shown to increase due to the finite size
effects. The distribution of mass and internal composition of the star for a
given value of the central pressure are shown in the right panel.  The dashed
lines show results for the case where finite size effects are included; this
is to be contrasted with the results obtained in the bulk approximation shown
by the solid curve. The spatial extent of the mixed phase region is also
influenced by finite size effects. The extent of the mixed phase region in the
inner core is shown for both cases and indicates that the finite size effects
cause the mixed phase region to shrink. It must however be noted that we are
comparing the structure of stars with different gravitational mass but the same
central pressure. 

\subsection{Transport Phenomena}
Transport properties in a mixed phase can differ greatly from those of uniform
matter since the mixed phase allows for new processes that cannot occur in a
translationally invariant system. Finite size effects play a central role in
the determination of the transport characteristics of a heterogenous phase
since they depend sensitively on the composition, size and geometry of the
structures present. In what follows we briefly discuss a few examples of
transport process which are strongly influenced by the finite size effects.

In a recent article, it was shown that low energy neutrinos couple coherently
to the droplet structures and consequently the neutrino mean free path in a
mixed phase was greatly reduced \cite{RBP}. The neutrino scattering rate
depends on the total weak charge, and on the filing fraction. The cross section
for the coherent neutrino-droplet scattering varies as the square of the total
weak charge. Thus, the neglect of screening effects alone, which could
overestimate the weak charge in a droplet by as much as $30-50$ \% (see
Fig.{\ref{screen}), could result in a mean free path estimate that was smaller
by $50-75$ \%. Furthermore, coulomb correlations between droplets decreases the
net neutrino scattering rate when the $E_\nu \lsim 1/d$, where $d=2R_{WS}$ is
the inter-droplet spacing.  Again, an accurate description of finite size and
Debye screening effects is required to calculate the neutrino mean free path in
the mixed phase \cite{RBP}. The filling fraction of the kaon phase, denoted by
$\chi$, at a pressure $p=0.40$ fm$^{-4}$ in the bulk approximation is
$\chi_{bulk}=0.23$ while the more detailed treatment yields $ \chi = 0.13$.
Again, this substantial change is due to the increase of the electron chemical
potential for small droplets.

The results of this work will also influence calculations of the melting
temperature of the lattice of droplets. This becomes obvious when we note that
the electric charge distributions are strongly modified. In addition, our
findings here affect the location and the nature of the interface between the
solid-mixed phase and the liquid-nuclear phase. The physics of this interface
has been shown to play an important role in damping r-mode instabilities in
cold and rapidly rotating neutron stars \cite{rmode}. In general, since the
transport characteristics depend primarily on the size and distance between
structures, this calls for a detailed description of the mixed phase wherein
screening and finite surface effects are incorporated. In particular, studies
of the early evolution of neutron stars \cite{PMPL}, where neutrino transport
plays an important will be strongly influenced by finite size effects discussed
here.
 
\section {Conclusions and Discussion}

The earliest papers on first-order kaon condensation in neutron stars ignored
finite size effects such as surface and Coulomb energies.  Since then, these
effects have been introduced into calculations by assuming that certain
quantities remain fixed as the physical size of geometric structures is varied.
By explicitly constructing these structures, however, we have shown these
assumptions to be unwarranted.  In particular, the energy densities, number
densities, and chemical potentials vary at the 10\% level as the size is
reduced toward the boundary thickness.  These changes in the bulk properties of
the two phases preclude the possibility of determining the most stable
structures by minimizing only the surface and Coulomb contributions to the
energy.  By including these effects correctly, we find that the critical
pressure for the onset of the mixed phase in increased dramatically, with the
further consequence that the maximum neutron star mass is increased back toward
its value for the pure nuclear equation of state.

Clearly, we have not accounted for all finite size effects that arise in the
mixed phase.  We examine some assumptions of our work here and comment on
corrections omitted.  The calculations presented here were based on the
Thomas-Fermi approximation for the nucleon fields, which we expect to be valid
based on our observation that the average nucleon density is large compared to
the size of the density gradients \cite{MS}. The Wigner-Seitz approximation
could in principle introduces corrections to the filling fraction. However, we
expect these corrections to be most important at large filling fraction where
the Wigner-Seitz description is accurate because the favored geometry is
planar.  In addition a careful treatment of the nucleon fields would introduce
shell effects that may be important for small structures.  These effects have
been neglected here since we expect their contributions to the Gibbs potential
to be small (typically, we expect that $\delta E_{shell} \sim A^{1/6}$ compared
to $A^{1}$ or $A^{2/3}$ dependence of volume and surface contributions, where
$A$ is the total baryon number of the droplet).

The increase in the critical pressure will also reduce the radial extent of the
mixed phase.  This will potentially affect our pictures of neutrino transport,
glitches, r-mode instabilities, and other issues involved in neutron star
structure and viscosity calculations.  Additionally, as pointed out by
Heiselberg, Pethick and Staubo \cite{HPS}, the mixed phase may not be realized
in practice if the nucleation time is too large, even if the mixed phase is
energetically favored.  Our work here gives a hint in that direction: correct
inclusion of surface and Coulomb energies dramatically reduces the energetic
favorability of the mixed phase over a wide range of pressures.  This increase
in energy, however, necessarily applies not only to the stable structures of
$5-6$ fm, but also to the smaller transitional structures which must be passed
through in the process of nucleation.  In short, our work suggests that the
energy barrier to nucleation may be somewhat higher than previously thought,
resulting in an increase of the expected time required for the phase transition
to occur.

The apparent experimental observation of relatively high-mass neutron stars
(e.g., Ref. \cite{ZSS,vanK,OK}) has been cited as evidence that kaon
condensation does not occur in neutron stars.  We hypothesize a different
interpretation which appears to be consistent with these high- mass
observations.  A sufficiently heavy proto-neutron star (PNS) may undergo a
kaon-condensate phase transition due to thermal nucleation while it is still
relatively hot.  This will of course soften the equation of state and perhaps
lead to gravitational collapse of a massive PNS to a black hole \cite{PREP}.
In contrast, a PNS which is insufficiently massive, on the other hand, will not
achieve the critical density for kaon-condensation and will therefore cool in
the standard way.  If it subsequently acquires mass through accretion the low
temperature may preclude the possibility of forming a kaon-condensed mixed
phase, even if the central pressure exceeds the critical pressure for the
transition.  The energy barrier to nucleation via quantum tunneling would
simply be too high, and the neutron star could exist in a metastable state for
a long time.  This (for now hypothetical) scenario would explain how the
observation of high-mass neutron stars (in excess of the maximum mass allowed
by the kaon mixed-phase equation of state) may not contradict the existence of
kaon condensation; these high-mass stars would need only to have acquired a
part of their mass through accretion after cooling, which appears in fact to be
likely. In any case, further examination of the time scale for nucleation of
kaon-condensate droplets is clearly interesting and warrants further work.

Finally, it should be pointed out that the relevance of these issues is not
restricted to the case of kaon condensation only.  Any first-order phase
transition that occurs in neutron stars will share the same qualitative
properties.  Hence, it is reasonable to suggest that a correct inclusion of
finite size effects in the deconfinement transition will have similar results
to the ones we have shown here, i.e., an increase in the critical pressure and
a possible increase in the nucleation time allowing for metastability.

\begin{center}

\large{ \textbf{Acknowledgements}}

\end{center}
We thank George Bertsch, Wick Haxton and Eduardo Fraga for useful discussions.
This work is supported in part by National Science Foundation Graduate Research
Fellowship (TN) and the US department of Energy grant DE-FG03-00-ER41132 (SR).

\newcommand{\IJMPA}[3]{{ Int.~J.~Mod.~Phys.} {\bf A#1}, (#2) #3}
\newcommand{\JPG}[3]{{ J.~Phys. G} {\bf {#1}}, (#2) #3}
\newcommand{\AP}[3]{{ Ann.~Phys. (NY)} {\bf {#1}}, (#2) #3}
\newcommand{\NPA}[3]{{ Nucl.~Phys.} {\bf A{#1}}, (#2) #3 }
\newcommand{\NPB}[3]{{ Nucl.~Phys.} {\bf B{#1}}, (#2)  #3 }
\newcommand{\PLB}[3]{{ Phys.~Lett.} {\bf {#1}B}, (#2) #3 }
\newcommand{\PRv}[3]{{ Phys.~Rev.} {\bf {#1}}, (#2) #3}
\newcommand{\PRC}[3]{{ Phys.~Rev. C} {\bf {#1}}, (#2) #3}
\newcommand{\PRD}[3]{{ Phys.~Rev. D} {\bf {#1}}, (#2) #3}
\newcommand{\PRL}[3]{{ Phys.~Rev.~Lett.} {\bf {#1}}, (#2) #3}
\newcommand{\PR}[3]{{ Phys.~Rep.} {\bf {#1}}, (#2) #3}
\newcommand{\ZPC}[3]{{ Z.~Phys. C} {\bf {#1}}, (#2) #3}
\newcommand{\ZPA}[3]{{ Z.~Phys. A} {\bf {#1}}, (#2) #3}
\newcommand{\JCP}[3]{{ J.~Comput.~Phys.} {\bf {#1}}, (#2) #3}
\newcommand{\HIP}[3]{{ Heavy Ion Physics} {\bf {#1}}, (#2) #3}
\newcommand{\RMP}[3]{{ Rev. Mod. Phys.} {\bf {#1}}, (#2) #3}
\newcommand{\APJ}[3]{{Astrophys. Jl.} {\bf {#1}}, (#2) #3}

\end{document}